\documentclass[11pt]{article}

\usepackage{amsmath}
\usepackage{graphicx,amssymb,amsbsy}

\def\pr{{\partial}}

\def\be{\mathbf{e}}

\def\bom{\boldsymbol{\omega}}
\newcommand\Real{\mbox{Re}}
\newcommand\Imag{\mbox{Im}}

\begin{document}

\centerline{\Large{\bf Inviscid instability of an incompressible flow}}
\centerline{\Large{\bf between rotating porous cylinders}}
\centerline{\Large{\bf to three-dimensional perturbations}}

\vskip 10mm
\centerline{\bf Konstantin Ilin\footnote{Department of Mathematics, University of York,
Heslington, York YO10 5DD, UK. Email: konstantin.ilin@york.ac.uk} and
Andrey Morgulis\footnote{Department of Mathematics, Mechanics and Computer Science, The Southern Federal University, Rostov-on-Don, and South Mathematical Institute, Vladikavkaz Center of RAS, Vladikavkaz, Russian Federation. Email: amor@math.sfedu.ru}}

\vskip 15mm
\begin{abstract}
We study the stability of two-dimensional inviscid flows in an annulus between two porous cylinders with respect to three-dimensional perturbations.
The basic flow is irrotational, and both radial and azimuthal
components of the velocity are non-zero. The direction of the radial flow can be from
the inner cylinder to the outer one (the diverging flow) or from the outer cylinder
to the inner one (the converging flow). It had been shown earlier in Ref. \cite{IM2013a} that, independent of the direction
of the radial flow, the basic flow can be unstable to small two-dimensional perturbations.
In the present paper, we prove first that purely radial flow is stable and that flows with
both radial and azimuthal components are always stable to axisymmetric perturbations. Then we show that both
the diverging and converging flows
are unstable with respect to non-axisymmetric three-dimensional perturbations provided that the ratio
of the azimuthal component of the velocity to the radial one is sufficiently large.
Neutral curves in the space of parameters of the problem are computed and it is demonstrated that for any ratio of the radii of the cylinders,
the most unstable modes (corresponding to the smallest ratio of the azimuthal velocity to the radial one) are the two-dimensional ones.
We also consider the corresponding viscous stability problem and construct an asymptotic expansion of its solutions for large radial
Reynolds numbers. We compute the first-order viscous correction to inviscid eigenvalues and show that
the asymptotic results give a good approximation to the viscous eigenvalues even for moderate values of radial Reynolds number, which
indicates that the instability may be observed in real flows.
\end{abstract}








\section{Introduction}
\label{sec:Intro}

In this paper we continue our study of the instability of a steady inviscid flow in an annulus between
two permeable rotating circular cylinders that had been started in Ref. \cite{IM2013a}. The
basic flow has both radial and azimuthal
components which are independent of the azimuthal angle $\theta$ and inversely proportional to the radial coordinate $r$ of the polar coordinates
system $(r, \theta)$ with the origin at the common axis of the cylinders. The direction of the radial flow can be from
the inner cylinder to the outer one (the diverging flow) or from the outer cylinder
to the inner one (the converging flow). It had been shown in Ref. \cite{IM2013a} that this flow can be unstable
to small two-dimensional perturbations (in the framework of the inviscid theory) provided that
the ratio of the azimuthal component of the velocity to its radial component is larger than
a certain critical value. This two-dimensional instability is oscillatory, and the neutral modes represent azimuthal travelling waves.
The main aim of the present study is to understand what happens if three-dimensional perturbations are allowed.
In particular, we are interested to investigate whether the most unstable mode (i.e. the mode that becomes unstable first when
the azimuthal component of the velocity is increased from $0$) is two-dimensional or not, and to determine the axial
wave number of the most unstable mode.

The stability of viscous flows between permeable rotating cylinders with a radial flow
had been studied by many authors \cite{Bahl, Chang, Min, Kolyshkin, Kolesov, Serre, Martinand}.
One of the main aims of these papers was to determine the effect of the radial flow
on the stability of the circular Couette-Taylor flow to axisymmetric perturbations, and the general conclusion was
that the radial flow affects the stability of the basic flow:
both a converging radial flow and a sufficiently strong diverging flow have a stabilizing effect on the Taylor instability,
but when a divergent flow is weak, it has a destabilizing effect \cite{Min, Kolyshkin}. However, it was not clear whether
a radial flow itself can induce instability for flows which are stable without it. This question had been answered affirmatively
by Fujita, Morimoto \& Okamoto \cite{Fujita} and later by Gallet, Doering and Spiegel \cite{Gallet2010} who had demonstrated that a particular classes of viscous flows between
porous rotating cylinders can be unstable to small two-dimensional perturbations.

Later it had been shown \cite{IM2013a} that both converging and diverging flows can be linearly
unstable in the framework of the inviscid theory and that the instability persists if small viscosity is taken into account.
In Ref. \cite{IM2013b}, a two-dimensional viscous stability problem had been considered, and it had been shown
that not only the particular classes of viscous steady flows considered in
\cite{Gallet2010, IM2013a} can be unstable to two-dimensional perturbations, but this is also true for a wide class
of steady rotationally-symmetric viscous flows
(without any restriction on angular velocities of the cylinders and for both converging and diverging flows).
A further development of the two-dimensional theory can be found in a recent paper by Kerswell \cite{Kerswell} where, among other things,
the effects of compressibility and nonlinearity have been considered.

In the inviscid theory, a purely azimuthal flow with the velocity inversely proportional to $r$ is stable not only to two-dimensional perturbations (see, e.g, \cite{Drazin}) but also to three-dimensional perturbations (this can be deduced from the sufficient condition for stability given
by Howard \& Gupta \cite{Howard}). In \cite{IM2013a}, it had shown that this stable flow becomes unstable to two-dimensional perturbations
if a radial flow is added and that the instability occurs for an arbitrarily weak radial flow (here `weak' means
`weak relative to the azimuthal flow').
Moreover, in the limit when the ratio of the radial component of the velocity to its azimuthal component
is small, the instability becomes independent of the only geometric parameter of the problem - the ratio of the radii of the cylinders.
It had also been observed that if the azimuthal component of the basic flow is zero, i.e. the flow is purely radial, then
it is stable (to two-dimensional perturbations). These facts indicate that the instability mechanism cannot be explained
in terms of known instabilities (e.g., such as shear flow instability or centrifugal instability). The asymptotic behaviour of the unstable
eigenmodes for weak radial flow shows that the limit when the radial component of the velocity
goes to zero is a singular limit of the linear stability problem \cite{IM2013a}. Adding a weak radial flow to a purely azimuthal one
results in formation of an \emph{inviscid} boundary layer near the inflow part of the boundary, and the new unstable
eigenmodes (that were absent in the purely azimuthal flow) appear within this boundary layer.

In the present paper, we examine the effect of three-dimension perturbations on the inviscid stability properties of the basic flow.
In particular, we rigorously prove that (i) the purely radial flow is stable to small
three-dimensional perturbations and (ii) the basic flow, in which both the radial and azimuthal components of the velocity are nonzero,
is always stable to axisymmetric perturbations. We also compute neutral curves on the plane of parameters of the problem, which demonstrate that,
the most unstable mode is always two-dimensional.

The outline of the paper is as follows. In Section 2, we introduce basic equations and formulate the linear stability problem.
Section 3 contains a linear inviscid stability analysis
of both the diverging and converging flows. In Section 4, the effect of viscosity is considered.
Discussion of the results is presented in
Section 5.


\section{Formulation of the problem}\label{sec:problem}

\subsection{Exact equations and basic steady flow}

We consider three-dimensional inviscid incompressible flows in the gap between two concentric circular cylinders
with radii $r_{1}$ and $r_{2}$ ($r_2 > r_1$). The cylinders are permeable for the fluid and there is a constant volume flux $2\pi Q$
(per unit length along the common axis of the cylinders) of
the fluid through the gap (the fluid is pumped into the gap at the inner cylinder and taken out at the outer one or {\em vice versa}).
$Q$ will be positive if the direction of the flow is from the inner cylinder to the outer one and negative if the flow direction is reversed.
Flows with positive and negative $Q$ will be referred to as diverging and converging flows respectively.
Suppose that
$r_1$ is taken as a length scale, $r^2_{1}/\vert Q\vert$ as a time scale, $\vert Q\vert/r_{1}$ as a scale for the velocity and $\rho Q^2/r_{1}^2$ for the pressure
where $\rho$ is the fluid density. Then the Euler equations, written in non-dimensional variables, have the form
\begin{eqnarray}
&&u_{t}+ u u_{r} + \frac{v}{r}u_{\theta} + w u_{z} -\frac{v^2}{r}= -p_{r} ,  \label{1} \\
&&v_{t}+ u v_{r} + \frac{v}{r}v_{\theta} + w u_{z} +\frac{u v}{r}= -\frac{1}{r} \, p_{\theta} ,  \label{2} \\
&&w_{t}+ u w_{r} + \frac{v}{r}w_{\theta} + w w_{z}  = - p_{z} ,  \label{3} \\
&&\frac{1}{r}\left(r u\right)_{r} +\frac{1}{r} \, v_{\theta} + w_{z}=0.  \label{4}
\end{eqnarray}
Here $(r,\theta,z)$ are the polar cylindrical coordinates, $u$, $v$ and $w$ are the radial, azimuthal and axial components
of the velocity and $p$ is the pressure.

It is well-known that if the flow domain is bounded by impermeable walls, then one needs to impose
the standard boundary condition of no normal velocity at the walls. This will guarantee that the the
resulting initial boundary value problem for the Euler equations is mathematically well-posed.
What is less known is that if there is a non-zero flow of the fluid through the boundary, then not only
the normal velocity must be given at the boundary, but some additional boundary conditions
at the inflow part of the boundary must also be imposed. What conditions should be added is a subtle question and there are
several answers that lead to mathematically correct initial boundary value problems (see, e.g., \cite{Monakh, MorgYud}). We
will use the boundary condition for the tangent component of the velocity, which at first approximation corresponds to
the condition at a porous cylinder (see \cite{Joseph}) and for which the corresponding mathematical problem is well-posed
(e.g., \cite{Monakh}).
Another important reason for using this condition is that it is consistent with the vanishing viscosity limit for the Navier-Stokes equations
(see, e.g., \cite{Temam, Ilin2008}). More precisely,
the solution of the Euler equations with a normal velocity prescribed on the entire boundary of the flow domain
and with a tangent velocity prescribed on the inflow part of the boundary represent the leading order term of
the asymptotic expansion of the solution of the corresponding viscous problem in which all components of the velocity are given on
the entire boundary.

So, our boundary conditions are
\begin{equation}
u\!\bigm\vert_{r=1}=\beta, \quad u\!\bigm\vert_{r=a}=\beta/a, \label{5a}
\end{equation}
where $a=r_2/r_1$ and $\beta=Q/\vert Q\vert$,
and either
\begin{equation}
\quad v\!\bigm\vert_{r=1}=\gamma,   \quad w\!\bigm\vert_{r=1}=0  \label{5b}
\end{equation}
for the diverging flow ($\beta=1$) or
\begin{equation}
\quad v\!\bigm\vert_{r=a}=\gamma/a,   \quad w\!\bigm\vert_{r=a}=0  \label{5c}
\end{equation}
for the converging flow ($\beta=-1$). In Eqs. (\ref{5b}) and (\ref{5c}),
$\gamma$ is the ratio of the azimuthal velocity to the radial velocity at the inner cylinder, i.e. $\gamma=\Omega_1 r_1^2/\vert Q\vert$ (where
$\Omega_1$ is angular velocity of the inner cylinder).

The problem, given by Eqs. (\ref{1})--(\ref{5a}) and (\ref{5b}) or (\ref{5c}),  has the following simple rotationally-symmetric solution:
\begin{equation}
u(r,\theta)=\beta/r, \quad v(r,\theta)=\gamma/r, \quad w=0.  \label{6}
\end{equation}
In what follows we will investigate the stability of this steady flow.

\subsection{Linear stability problem}

Consider a small perturbation
$(\tilde{u}, \tilde{v}, \tilde{w}, \tilde{p})$ of the basic flow (\ref{6}). It is convenient to write the linearised equations in the terms of perturbation vorticity
\begin{equation}
\bom =\omega_1 \, \be_r + \omega_2 \, \be_{\theta} + \omega_3 \, \be_z   \label{7}
\end{equation}
where $\be_r$, $\be_{\theta}$ and $\be_z$ are unit vectors in the radial, azimuthal and axial directions, respectively,
and where
\begin{eqnarray}
&&\omega_1 = \frac{1}{r} \, \tilde{w}_{\theta} - \tilde{v}_{z},  \label{8} \\
&&\omega_2 = \tilde{u}_{z} - \tilde{w}_{r},  \label{9} \\
&&\omega_3 = \frac{1}{r} \, \left( (r \tilde{v})_{r} - \tilde{u}_{\theta}\right). \label{10}
\end{eqnarray}
The linearised equation can be written as
\begin{eqnarray}
&&\left(\pr_{t} + \frac{\gamma}{r^2} \, \pr_{\theta}+\frac{\beta}{r} \, \pr_{r}+\frac{\beta}{r^2}\right)\omega_1 =0,  \label{11} \\
&&\left(\pr_{t} + \frac{\gamma}{r^2} \, \pr_{\theta}+\frac{\beta}{r} \, \pr_{r}-\frac{\beta}{r^2}\right)\omega_2 =
-\frac{2\gamma}{r^2} \, \omega_1,  \label{12} \\
&&\left(\pr_{t} + \frac{\gamma}{r^2} \, \pr_{\theta}+\frac{\beta}{r} \, \pr_{r}\right)\omega_3 =0. \label{13}
\end{eqnarray}
We seek a solution of Eqs. (\ref{8})--(\ref{13}) in the form of the normal mode
\[
\{\tilde{u}, \tilde{v}, \tilde{w}, \bom \} =
Re\left[\{\hat{u}(r), \hat{v}(r), \hat{w}(r), \hat{\bom}(r) \}
e^{\sigma t + in\theta+ikz}\right]
\]
where $n\in\mathbb{Z}$ and $k\in\mathbb{R}$ are the azimuthal and axial wave numbers respectively.
On substituting this into Eqs. (\ref{11})--(\ref{13}), we can rewrite them as
\begin{eqnarray}
&&\left(h(r)+\frac{\beta}{r} \, \pr_{r} \right) (r\hat{\omega}_1) =0,  \label{14} \\
&&\left(h(r)+\frac{\beta}{r} \, \pr_{r} \right) \left(\frac{\hat{\omega}_2}{r}\right) =
-\frac{2\gamma}{r^3} \, \omega_1,  \label{15} \\
&&\left(h(r)+\frac{\beta}{r} \, \pr_{r} \right)\hat{\omega}_3 =0 \label{16}
\end{eqnarray}
where
\begin{eqnarray}
&&h(r) = \sigma +\frac{in\gamma}{r^2},  \label{17} \\
&&\hat{\omega}_1 = \frac{in}{r}\, \hat{w} - ik \hat{v},  \label{18} \\
&&\hat{\omega}_2 = ik \hat{u} - \hat{w}_r,  \label{19} \\
&&\hat{\omega}_3 = \frac{1}{r} \, \left( r \hat{v}\right)_{r} - \frac{in}{r}\, \hat{u}.  \label{20}
\end{eqnarray}
Equations (\ref{14})--(\ref{16}) should be solved subject to the boundary conditions:
\begin{eqnarray}
&&\hat{u}\!\bigm\vert_{r=1}=0,  \label{21} \\
&&\hat{u}\!\bigm\vert_{r=a}=0. \label{22}
\end{eqnarray}
and either
\begin{equation}
\hat{v}\!\bigm\vert_{r=1}=0,  \quad
\hat{w}\!\bigm\vert_{r=1}=0 \label{23}
\end{equation}
for the diverging flow or
\begin{equation}
\hat{v}\!\bigm\vert_{r=a}=0,  \quad
\hat{w}\!\bigm\vert_{r=a}=0 \label{24}
\end{equation}
for the converging flow.
Equations (\ref{14})--(\ref{22}) together with either (\ref{23}) or (\ref{24}) represent an eigenvalue problem for $\sigma$. If there is an eigenvalue $\sigma$ such that
$Re(\sigma)>0$, then the basic flow is unstable. If there are no eigenvalues with positive real part and if there are no
perturbations with non-exponential growth (examples of non-exponential growth can be found, e.g., in \cite{Barlow}), then the flow is linearly stable. In the next section we analyse this eigenvalue problem.


\section{Analysis of the eigenvalue problem}

The eigenvalue problem formulated above can be reduced to a problem of finding zeros of a certain entire function.
We will show this first for the divergent flow.

\subsection{Diverging flow ($\beta=1$)}

\subsubsection{Dispersion relation}

Boundary conditions (\ref{23}) and Eq. (\ref{18}) imply that
\begin{equation}
\hat{\omega}_1\!\bigm\vert_{r=1}=0. \label{3.1}
\end{equation}
Now let
\begin{equation}
g(r)=\sigma \, \frac{r^2}{2} + in\gamma \, \ln r, \label{3.2}
\end{equation}
so that $h(r)$, given by (\ref{17}), can be written as $h(r)=g'(r)/r$.
Then the general solution of Eq. (\ref{14}) is
\[
r\hat{\omega}_1 = C e^{-g(r)}
\]
where $C$ is an arbitrary constant. This and Eq. (\ref{3.1}) imply that
$C=0$ and, therefore, $\hat{\omega}_1(r)=0$, so that we have the relation
\begin{equation}
\frac{in}{r}\, \hat{w} - ik \hat{v}=0. \label{3.3}
\end{equation}
Now we assume that $n\neq 0$. The case of $n=0$ will be treated separately.
Using (\ref{3.3}) to eliminate $\hat{w}$ from the incompressibility condition
\begin{equation}
\frac{1}{r}\, (r \hat{u})_{r} + \frac{in}{r}\, \hat{v} + ik\hat{w}=0, \label{3.4}
\end{equation}
we obtain
\begin{equation}
\frac{in}{r}\, (r \hat{u})_{r} -\left(k^2 + \frac{n^2}{r^2}\right) r\hat{v}=0. \label{3.5}
\end{equation}
Integration of Eq. (\ref{16}) yields
\begin{equation}
\hat{\omega}_3 = C_1 \, e^{-g(r)}  \label{3.6}
\end{equation}
for an arbitrary constant $C_1$. Equations (\ref{3.6}) and (\ref{20}) have a consequence that
\begin{equation}
inr \hat{u} = r(r\hat{v})_r - C_1 r^2 e^{-g(r)}.  \label{3.7}
\end{equation}
Finally, we use (\ref{3.7}) to eliminate $\hat{u}$ from Eq. (\ref{3.5}). As a result, we get the equation
\begin{equation}
G_{rr} + \frac{1}{r}\, G_r - \left(k^2 + \frac{n^2}{r^2}\right) G  = C_1 \, F(r)  \label{3.8}
\end{equation}
where
\begin{equation}
G(r)  = r\hat{v}(r)  \label{3.9}
\end{equation}
and
\begin{equation}
F(r)  = \frac{1}{r} \, \pr_{r}\left(r^2 e^{-g(r)}\right).  \label{3.10}
\end{equation}
Equation (\ref{3.7}) allows us to rewrite boundary conditions (\ref{21})--(\ref{23}) (for $\hat{u}$ and $\hat{v}$)
in terms of $G$:
\begin{eqnarray}
&&G(1)=0,  \label{3.11} \\
&&G'(1)=C_1 \, e^{-g(1)},  \label{3.12} \\
&&G'(a)=C_1 \, a \,  e^{-g(a)}. \label{3.13}
\end{eqnarray}
Equation (\ref{3.8}) together with boundary conditions (\ref{3.11})--(\ref{3.13}) represent an eigenvalue problem for $\sigma$ (that
enters the problem via $g(r)$).

The general solution of Eq. (\ref{3.8}) can be written as
\begin{equation}
G(r)  = C_1 \int\limits_{1}^{r} F(s)\left[I_n(kr)K_n(ks)- I_n(ks)K_n(kr)\right] s \, ds +
C_2 \, I_n(kr) + C_3 \, K_n(kr).  \label{3.14}
\end{equation}
Here $I_n(z)$ and $K_n(z)$ are the modified Bessel functions of the first and second kind; $C_2$ and $C_3$ are arbitrary constants (recall that $C_1$ is also arbitrary).
Substitution of the general solution into boundary conditions (\ref{3.11}) and (\ref{3.12}) results in the following two equations:
\begin{eqnarray}
&&C_2 \, I_n(k) + C_3 \, K_n(k)=0,  \nonumber \\
&&C_2 \, k I'_n(k) + C_3 \, k K'_n(k)=C_1  e^{-g(1)}.  \nonumber
\end{eqnarray}
Solving these for $C_1$ and $C_2$, we obtain
\begin{equation}
C_2 = C_1 \,  K_n(k)  e^{-g(1)}, \quad C_3 = - C_1 \,  I_n(k)  e^{-g(1)}.  \label{3.15}
\end{equation}
Here we have used the Wronskian relation (e.g. \cite{Abramowitz}):
\begin{equation}
I_{n}'(z) K_{n}(z)-I_{n}(z) K_{n}'(z)=\frac{1}{z}.  \label{3.16}
\end{equation}
With the help of (\ref{3.15}), we can rewrite Eq. (\ref{3.14}) in the form
\[
G(r)  = C_1 \left\{\int\limits_{1}^{r} F(s)\left[I_n(kr)K_n(ks)- I_n(ks)K_n(kr)\right] s \, ds +
\left[I_n(kr)K_n(k) - I_n(k)K_n(kr)\right] e^{-g(1)}\right\}.
\]
Substituting this into boundary condition (\ref{3.13}), we obtain the dispersion relation
\begin{eqnarray}
&&\int\limits_{1}^{a} F(s)k\left[ I'_n(ka)K_n(ks)- I_n(ks) K'_n(ka)\right] s \, ds \nonumber \\
&&\qquad\qquad\qquad +
k\left[ I'_n(ka)K_n(k) - I_n(k) K'_n(ka)\right] e^{-g(1)}-  a \,  e^{-g(a)} =0. \label{3.17}
\end{eqnarray}
This dispersion relation can be further simplified as follows. Let $\mathcal{I}$ be the integral entering the dispersion relation.
Recalling that $F(r)$ is given by Eq. (\ref{3.10}) and integrating by parts, we obtain
\begin{eqnarray}
\mathcal{I}&=&k\int\limits_{1}^{a} \frac{1}{s} \, \pr_{s}\left(s^2 e^{-g(s)}\right)\left[I'_n(ka)K_n(ks)- I_n(ks) K'_n(ka)\right] s \, ds
\nonumber \\
&=&\left. s^2 e^{-g(s)}k \left[ I'_n(ka)K_n(ks)- I_n(ks) K'_n(ka)\right]\right\vert_{1}^{a} \nonumber \\
&& -k^2 \int\limits_{1}^{a} e^{-g(s)}\left[ I'_n(ka)K'_n(ks)- I'_n(ks) K'_n(ka)\right] s^2 \, ds \nonumber \\
&=&a e^{-g(a)}- e^{-g(1)} k \left[ I'_n(ka)K_n(k)- I_n(k) K'_n(ka)\right] \nonumber \\
&& -k^2 \int\limits_{1}^{a} e^{-g(s)}\left[ I'_n(ka)K'_n(ks)- I'_n(ks) K'_n(ka)\right] s^2 \, ds .  \nonumber
\end{eqnarray}
Here again we have used the Wronskian relation (\ref{3.16}).
Substitution of the above formula for $\mathcal{I}$ into (\ref{3.17}) yields the final expression for the dispersion relation:
\begin{equation}
D(\sigma,n,k,\gamma,a)\equiv k^2 \int\limits_{1}^{a} e^{-\sigma s^2/2-in\gamma\ln s}\left[ I'_n(ks)K'_n(ka)- I'_n(ka) K'_n(ks)\right] s^2 \, ds
=0 . \label{3.18}
\end{equation}
It can be shown that in the limit $k\to 0$ this reduces to the dispersion relation of the corresponding two-dimensional problem (considered in \cite{IM2013a}).

The dispersion relation (\ref{3.18}) has been obtained under assumption that $n\neq 0$. Nevertheless, it can be shown that this dispersion relation is also valid for the axisymmetric mode, $n=0$.

The eigenfunction $G(r)$ associated with the eigenvalue $\sigma$ can be written as
\[
G(r)= C_1 \, k \int\limits_{1}^{r} e^{-\sigma s^2/2-in\gamma\ln s}\left[ I'_n(ks)K_n(kr)- I_n(kr) K'_n(ks)\right] s^2 \, ds ,
\]
while the corresponding formula for $H(r)\equiv r\hat{u}(r)$ is
\[
H(r)= C_1 \, \frac{k^2}{in} \,  r \int\limits_{1}^{r} e^{-\sigma s^2/2-in\gamma\ln s}\left[ I'_n(ks)K'_n(kr)- I'_n(kr) K'_n(ks)\right] s^2 \, ds .
\]


\subsubsection{General properties of the dispersion relations (\ref{3.18})}

It has been mentioned in \cite{IM2014} that certain conclusions about a two-dimensional counterpart of (\ref{3.18}) can be made using
the P\'{o}lya theorem (see problem 177 of Part V in \cite{Polya_book}, see also \cite{Polya1918}). It turns out that this theorem also works
for (\ref{3.18}). It is shown in Appendix A that, for the purely radial flow ($\gamma=0$), the dispersion relation
(\ref{3.18}) has no roots with non-negative real part, so that there are no growing normal modes for the purely radial diverging flow.
The same is true for the axisymmetric mode, $n=0$ (see Appendix A). So, we can restrict our attention
to non-axisymmetric perturbations for $\gamma\neq 0$.

Also, using the fact that $I_{-n}(z)=I_{n}(z)$ and $K_{-n}(z)=K_{n}(z)$ (e.g. \cite{Abramowitz}), we deduce from (\ref{3.18}) that
\begin{eqnarray}
&&\overline{D(\sigma, n, k, a, \gamma)}=D(\bar{\sigma}, -n, k, a, \gamma), \label{3.19} \\
&&D(\sigma, n, k, a, \gamma)=D(\sigma, -n, k, a, -\gamma) \label{3.20}
\end{eqnarray}
where the bar denotes complex conjugation. These relations imply that
it suffices to consider
only positive $n$ and $\gamma$.

\subsubsection{Numerical results}

\begin{figure}
\begin{center}
\includegraphics*[height=7cm]{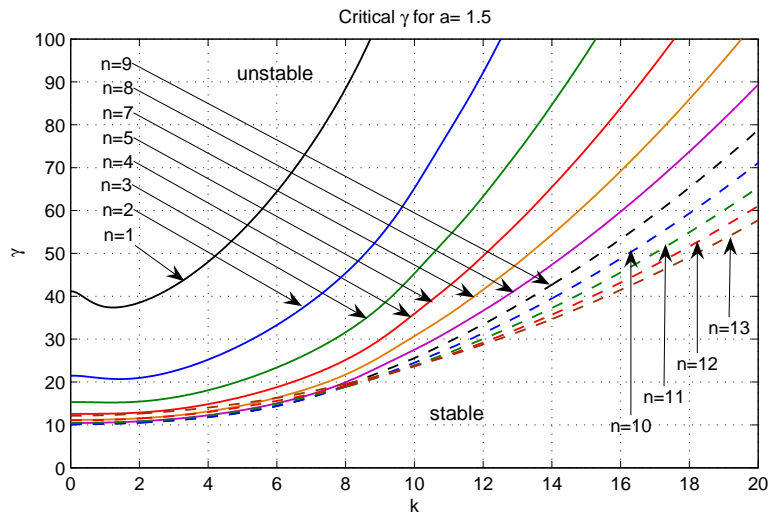}
\end{center}
\caption{Neutral curves for $\beta=1$ (diverging flow), $a=1.5$ and $n=1,\dots,13$. The region above each curve is where the corresponding mode is unstable.}
\label{cr_gamma}
\end{figure}

\begin{figure}
\begin{center}
\includegraphics*[height=7cm]{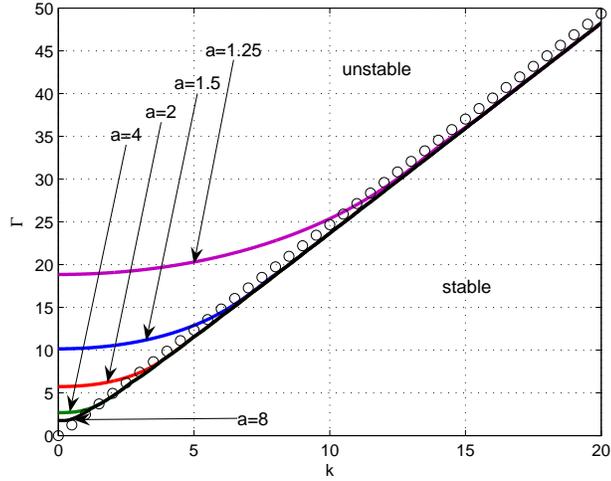}
\end{center}
\caption{Critical $\gamma$ minimized over azimuthal wave numbers $n=1,2, \dots 20$ ($\Gamma=\min_{n}\gamma_{cr}$) for the diverging flow and
$a=1.25, 1.5, 2, 4, 8$. The region above each curve is where the corresponding flow is unstable. Circles show the asymptotic stability
boundary for large $k$ derived in Appendix B.}
\label{min_cr_gamma}
\end{figure}

As we already know, for $\gamma=0$, all eigenvalues lie in the left half-plane of complex variable $\sigma$. Numerical evaluation of (\ref{3.18}) confirms this fact and shows that when $\gamma$ increases from $0$, some eigenvalues move to the right, and
there is a critical value $\gamma_{cr}>0$ of parameter $\gamma$ at which one of the eigenvalues crosses the imaginary axis, so that
\[
\Real (\sigma) > 0 \ \ {\rm for} \ \ \gamma >\gamma_{cr} \  \ {\rm and} \ \ \Real (\sigma) < 0 \ \
{\rm for} \ \ \gamma <\gamma_{cr}.
\]
We have computed neutral curves ($\Real (\sigma) = 0$) on the $(k,\gamma)$ plane for several values of the geometric parameter $a$ and for $n=1,2, \dots 20$. For all $a$, the neutral curves look qualitatively similar to what is shown in Fig. \ref{cr_gamma}. One can see that
the neutral curves for a few modes with low azimuthal wave number can be non-monotonic functions of the axial wave number $k$ (e.g., $n=1,2,3$ in Fig. \ref{cr_gamma}). However, all other
modes are strictly increasing functions of $k$. Let $\Gamma (k)$ be the critical value of $\gamma$ minimized over $n=1, \dots,20$:
\[
\Gamma (k)= \min_{n} \gamma_{cr}(n,k).
\]
Functions $\Gamma (k)$ for several values of the geometric parameter $a$ are shown in Fig. \ref{min_cr_gamma}.
This figure demonstrates the following three things. First,
$\Gamma (k)$, for any value of the geometric parameter $a$, is an increasing function,
so that its minimum is attained at $k=0$, i.e. for the two-dimensional mode. Thus the mode that
becomes unstable first when $\gamma$ increases from $0$ (we will call it the most unstable mode) is two-dimensional.
Second, for small to moderate values of $k$ ($k\lesssim 10$), function $\Gamma(k)$ considerably depends on $a$:
on one hand, it decreases when $a$ is increased and seems to tend to a limit for large $a$; on the other hand, it grows
when $a$ tends to $1$.
Third,
$\Gamma (k)$, for any value of $a$, becomes a linear function of $k$ for sufficiently large $k$. Moreover, this linear asymptote
is the same for all values of $a$. It is shown in Appendix B that in the limit of large $k$ and $n$, more precisely, if
\[
n=\alpha k \quad \hbox{and} \quad k\to\infty,
\]
where $\alpha$ is a positive constant, then
\[
\min_{\alpha}\gamma_{cr}(\alpha, k)\sim  2.4671 \, k .
\]
This asymptotic result is shown by circles in Fig. \ref{min_cr_gamma}. Evidently, it is in a good agreement with the numerical results.
\begin{figure}
\begin{center}
\includegraphics*[height=7cm]{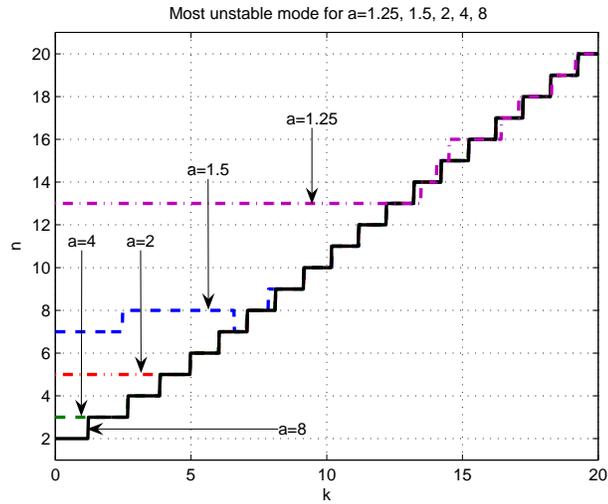}
\end{center}
\caption{The azimuthal wave number of the most unstable mode, $n$, for the diverging flow as a function of $k$ for $a=1.25, 1.5, 2, 4, 8$.}
\label{most_unstable_mode}
\end{figure}
The azimuthal wave number $n$ of the most unstable mode (that, for a fixed $k$, becomes unstable first when $\gamma$ is increased from $0$)
depends on both $a$ and $k$. The results of the numerical calculations of this quantity are shown in Fig. \ref{most_unstable_mode}.
The jumps in $n$ correspond to the intersection points of the neutral curves for individual azimuthal modes. Figure \ref{most_unstable_mode}
indicates that, for sufficiently large $k$, the azimuthal wave number of the most unstable mode, $n$, is independent of $a$ and $n\sim k$.
These facts are employed in Appendix B where the asymptotic behaviour of eigenvalues for large $k$ is considered.

\begin{figure}
\begin{center}
\includegraphics*[height=7cm]{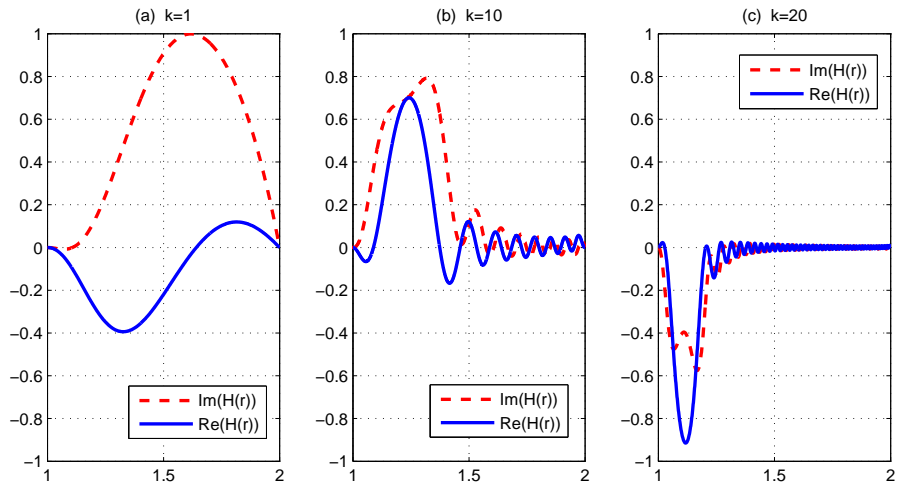}
\end{center}
\caption{Eigenfunction $H(r)$ for neutral modes ($\gamma=\gamma_{cr}$) for the diverging flow, $a=2$ and $n=4$: (a) $k=1$, $\gamma=5.9483$;
(b) $k=10$, $\gamma=35.7645$; (c)  $k=20$, $\gamma=128.2941$.}
\label{eigenfunctions}
\end{figure}

The graphs of functions $\Real\left(H(r)\right)$ and $\Imag\left(H(r)\right)$ corresponding to the critical value of $\gamma$
for $a=2$ and $n=4$ are shown in Fig. \ref{eigenfunctions}. Evidently, when the axial wave number increases, the eigenfunction becomes
more oscillatory and concentrated near the inner cylinder (i.e. at the flow inlet). This fact is also used in the investigation of
the asymptotic behaviour of eigenvalues for large $k$ in Appendix B.

\subsection{Converging flow ($\beta=-1$)}

\subsubsection{Dispersion relation}

An analysis similar to what we did for $\beta=1$ results in the following dispersion relation
\begin{equation}
D_1(\sigma,n,k,\gamma,a)\equiv k^2 \int\limits_{1}^{a} e^{\sigma s^2/2+in\gamma\ln s}\left[ I'_n(ks)K'_n(k)- I'_n(k) K'_n(ks)\right] s^2 \, ds
=0 . \label{3.30}
\end{equation}
Again, it can be shown that in the limit $k\to 0$ this reduces to the dispersion relation of the corresponding two-dimensional problem (see \cite{IM2013a}).

Similarly to how this was done in Appendix B for the diverging flow, it can be shown that the dispersion relation
(\ref{3.30}) has no roots with non-negative real part (i) for the purely radial converging flow
(i.e. for $\gamma=0$ and for all $n$) and (ii) for the axisymmetric mode (for $n=0$ and for all $\gamma$).

The dispersion relation (\ref{3.30}) has the same symmetry properties as its counterpart (\ref{3.18}) for the diverging flow:
\begin{eqnarray}
&&\overline{D_1(\sigma, n, k, a, \gamma)}=D_1(\bar{\sigma}, -n, k, a, \gamma), \label{3.31} \\
&&D_1(\sigma, n, k, a, \gamma)=D_1(\sigma, -n, k, a, -\gamma). \label{3.32}
\end{eqnarray}
These relations imply that
we need to consider only positive $n$ and $\gamma$.

\subsubsection{Numerical results}

\begin{figure}
\begin{center}
\includegraphics*[height=7cm]{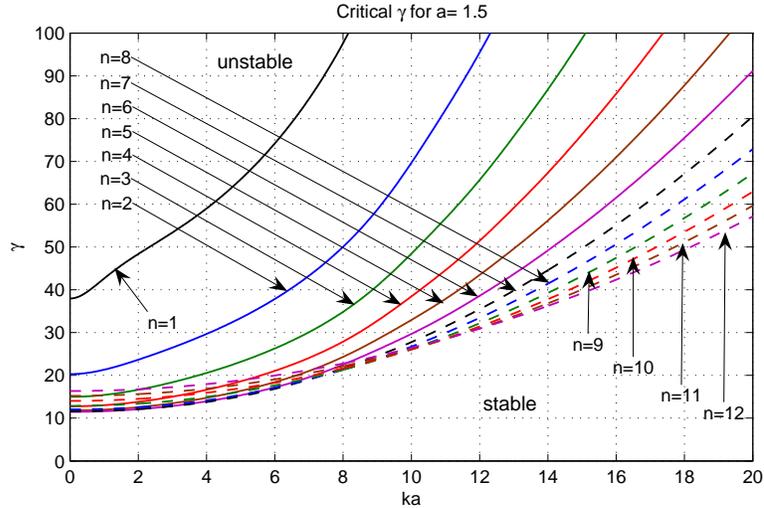}
\end{center}
\caption{Neutral curves for $\beta=-1$ (converging flow), $a=1.5$ and $n=1,\dots,12$.
The region above each curve is where the corresponding mode is unstable.}
\label{cr_gamma_conv}
\end{figure}

\begin{figure}
\begin{center}
\includegraphics*[height=7cm]{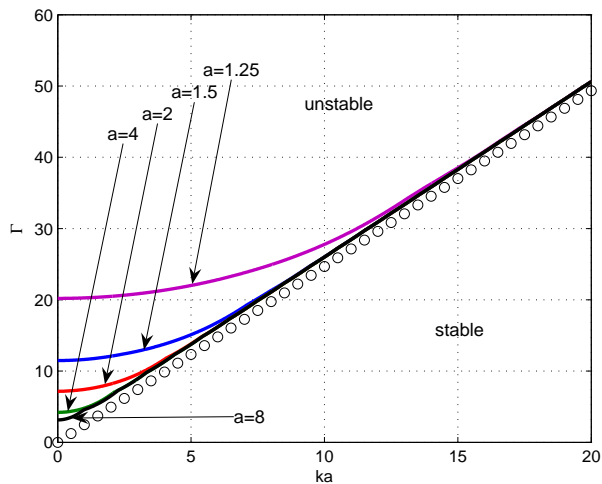}
\end{center}
\caption{Critical $\gamma$ minimized over azimuthal wave numbers $n=1,2, \dots 20$ ($\Gamma=\min_{n}\gamma_{cr}$) for the converging flow and
$a=1.25, 1.5, 2, 4, 8$. The region above each curve is where the corresponding flow is unstable. Circles represent the asymptotic
stability boundary for large $ka$ derived in Appendix B.}
\label{min_cr_gamma_conv}
\end{figure}

\begin{figure}
\begin{center}
\includegraphics*[height=7cm]{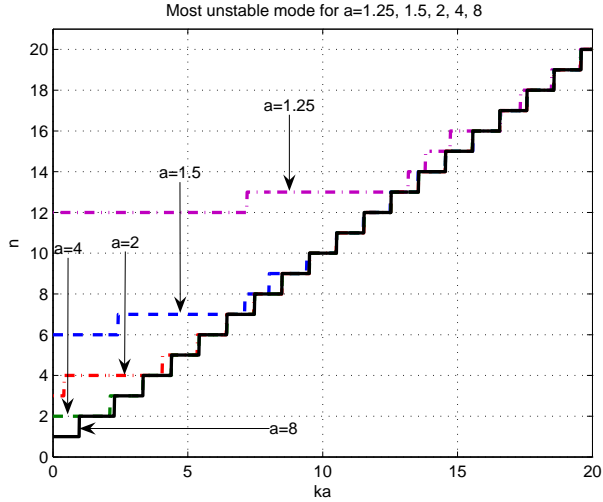}
\end{center}
\caption{The azimuthal wave number of the most unstable mode, $n$, for the converging flow versus $ka$ for $a=1.25, 1.5, 2, 4, 8$.}
\label{most_unstable_mode_conv}
\end{figure}

Numerical results for the converging flow are similar to those for the diverging flow: for $\gamma=0$, all eigenvalues lie in the left half-plane of complex variable $\sigma$; when $\gamma$ increases from $0$, some eigenvalues move to the right and cross the imaginary axis.
In the case of the converging flow, we will use parameter $ka$ instead of $k$. This is convenient because, to a certain extent,
it allows us to eliminate the dependence of the results on the geometric parameter $a$.
We have computed neutral curves ($\Real (\sigma) = 0$) on the $(ka,\gamma)$ plane for several values of the geometric parameter
$a$ and for $n=1,2, \dots 20$. For all $a$, the neutral curves look qualitatively similar to what is shown for $a=1.5$ in Fig. \ref{cr_gamma_conv}. We have found that, at least for $a=1.25, 1.5, 2, 4$ and $8$, the neutral curves for all azimuthal modes are increasing functions of $k$
(this differs from the case of the diverging flow where neutral curves for some low azimuthal modes can have a local minimum,
e.g. for the modes with $n=1,2$ in Fig. \ref{cr_gamma}).
Let $\Gamma (ka)$ be the critical value of $\gamma$ minimized over $n=1,2, \dots 20$:
\[
\Gamma (ka)= \min_{n} \gamma_{cr}(n,ka).
\]
Functions $\Gamma (ka)$ for several values of the geometric parameter $a$ are shown in Fig. \ref{min_cr_gamma_conv}.
The following conclusions can be drawn from this figure. First,
$\Gamma (ka)$ is an increasing function for any value of the geometric parameter $a$ (at least in the range $1.25\leq a\leq 8$),
so that its minimum is attained at $k=0$, i.e. for the two-dimensional mode. So, the mode that
becomes unstable first when $\gamma$ increases from $0$ is two-dimensional.
Second, one can see that, for small to moderate values of $ka$ ($ka\lesssim 10$), the critical value of $\gamma$ depends on $a$:
it decreases when $a$ is increased and seems to tend to a limit for large $a$; and it grows
when $a$ decreases.
Third, for any values of $a$, $\Gamma (ka)$ becomes a linear function for sufficiently large $ka$, and this linear asymptote
is the same for all values of $a$. We show in Appendix B that if
\[
n=\alpha \, ka \quad \hbox{as} \quad ka\to\infty,
\]
where $\alpha$ is a positive constant, then
\[
\min_{\alpha}\gamma_{cr}(\alpha, ka)\sim  2.4671 \, ka .
\]
This asymptotic result is shown by circles in Fig. \ref{min_cr_gamma_conv}. One can see that it is in a good agreement with the numerical results
even if $ka$ is not very large. The comparison of Figures \ref{min_cr_gamma} and \ref{min_cr_gamma_conv} shows that the critical values of
$\gamma$ for the converging flow is slightly higher than that for the diverging flow. In this sense, the converging flow is more stable than the
diverging flow.
The azimuthal wave number $n$ of the most unstable mode (that, for a fixed $ka$, becomes unstable first when $\gamma$ is increased from $0$)
depends on both $a$ and $ka$. The results of the numerical calculations are shown in Fig. \ref{most_unstable_mode_conv}.
The jumps in $n$ correspond to the intersection points of the neutral curves for individual azimuthal modes. One can see
in Fig. \ref{most_unstable_mode_conv}
that, for sufficiently large $ka$, the azimuthal wave number of the most unstable mode, $n$, is independent of $a$ and $n\sim ka$.
These facts are used in Appendix B.


\section{Effect of viscosity}

Here our aim is to show that for sufficiently high Reynolds numbers the unstable inviscid modes
found in the previous section give a good approximation to the corresponding viscous modes. We will restrict our analysis to
the diverging flow and compute the first viscous correction to the inviscid eigenvalues in the limit of high radial Reynolds numbers.

The most general steady rotationally-symmetric {\em viscous}
flow between rotating porous cylinders is given by (see, e.g., \cite{IM2013b})
\begin{equation}
u=\frac{\beta}{r}, \quad v=V(r)=A \, r^{1+\beta R} + \frac{B}{r}, \label{4.1}
\end{equation}
where $R=\vert Q\vert/\nu$ is the radial Reynolds number, (with $\nu$ being the kinematic viscosity), and
$A$ and $B$ are constants which depend on
$a$, $R$, $\gamma_1=\Omega_1 r_1^2/Q$ and $\gamma_2=\Omega_2 r_2^2/Q$ (with $\Omega_1$ and $\Omega_2$ being the angular velocities
of the inner and outer cylinders). We will consider only the diverging flow, i.e.
$\beta=1$ in Eq. (\ref{4.1}).

It can be shown (see, e.g., \cite{IM2013b}) that
in the limit
of high Reynolds numbers, $R\gg 1$, the azimuthal component of the velocity is well approximated by
\begin{equation}
V(r)=\frac{\gamma_1}{r}+\frac{f(\eta)}{a} + O\left(R^{-1}\right) \label{4.2}
\end{equation}
where $\eta=R(1-r/a)$ is the boundary layer variable and function $f$ is defined as
\begin{equation}
f(z)=(\gamma_2 - \gamma_1)e^{-z}. \label{4.3}
\end{equation}
Note that the second term in (\ref{4.2}) (the boundary layer term) is non-zero only
in the layer of thickness $O(R^{-1})$ near the outer cylinder, so that the flow is well approximated by the first term everywhere except for this thin boundary layer. The first term represents the inviscid flow that coincides with (\ref{6}) (up to replacing $\gamma_1$ with $\gamma$).
Note also that the single inviscid flow (\ref{6}) represents the high-Reynolds-number limit of each member
of a one-parameter family of viscous flows (parametrised by $\gamma_2$).

Consider a small perturbation
$(\tilde{u}, \tilde{v}, \tilde{w}, \tilde{p})$ in the form of the normal mode
\[
\{\tilde{u}, \tilde{v}, \tilde{w}, \tilde{p}\} = Re\left[\{\hat{u}(r), \hat{v}(r), \hat{w}(r),
\hat{p}(r)\} e^{\sigma t + in\theta+ikz}\right]
\]
where $\tilde{u}$, $\tilde{v}$, $\tilde{w}$ are perturbations of the radial, azimuthal and axial components of
the velocity, $\tilde{p}$ is the perturbation pressure, $n\in\mathbb{Z}$ and $k\in\mathbb{R}$. Substituting this into the linearised Navier-Stokes equations yields the eigenvalue problem:
\begin{eqnarray}
&&\left(\sigma +  \frac{in V}{r} + \frac{\beta}{r} \, \pr_r \right) \hat{u}
-\frac{\beta}{r^2} \, \hat{u} -\frac{2V}{r} \, \hat{v} = - \pr_r \, \hat{p}  +
\frac{1}{R} \left(L \hat{u}-\frac{\hat{u}}{r^2}-\frac{2in}{r^2} \, \hat{v}\right) ,  \label{4.4} \\
&&\left(\sigma +  \frac{in V}{r} + \frac{\beta}{r} \, \pr_r \right) \hat{v}
+\frac{\beta}{r^2} \, \hat{v}  +\Omega(r) u = -\frac{in}{r} \, \hat{p}  +
\frac{1}{R} \left(L \hat{v} -\frac{\hat{v}}{r^2}+\frac{2in}{r^2} \, \hat{u}\right),  \label{4.5} \\
&&\left(\sigma +  \frac{in V}{r} + \frac{\beta}{r} \, \pr_r \right) \hat{w} = -i k \, \hat{p}  +
\frac{1}{R} \, L \hat{w},  \label{4.6} \\
&&\pr_r \left(r \hat{u}\right) +in \, \hat{v} + ik r \, \hat{w}=0, \label{4.7} \\
&&\hat{u}(1)=0, \quad \hat{u}(a)=0, \quad \hat{v}(1)=0, \quad \hat{v}(a), \quad \hat{w}(1)=0, \quad \hat{w}(a)=0.  \label{4.8}
\end{eqnarray}
In Eqs. (\ref{4.4})--(\ref{4.6}),
\[
L  = \frac{d^2}{dr^2} + \frac{1}{r} \, \frac{d}{dr} - \left(k^2 + \frac{n^2}{r^2}\right), \quad \Omega(r)=V'(r)+\frac{V(r)}{r} .
\]
For $k=0$, this system reduces to the two-dimensional viscous stability problem whose asymptotic behaviour has been studied
in \cite{IM2013b}. An asymptotic expansion of solutions of Eqs. (\ref{4.4})--(\ref{4.8}) can be constructed in almost exactly
the same manner and has the form
\begin{eqnarray}
&& \sigma=\sigma_{0}+R^{-1}\sigma_{1}+O\left(R^{-2}\right),  \label{4.9} \\
&& \hat{u}=\hat{u}^{r}_{0}(r)+R^{-1} [\hat{u}^{r}_{1}(r)+\hat{u}^{b}_{0}(\eta)]+O\left(R^{-2}\right)  \label{4.10} \\
&& \hat{v}=\hat{v}^{r}_{0}(r)+\hat{v}^{b}_{0}(\eta)+ R^{-1} [\hat{v}^{r}_{1}(r)+\hat{v}^{b}_{1}(\eta)]+O\left(R^{-2}\right)  \label{4.11} \\
&& \hat{w}=\hat{w}^{r}_{0}(r)+\hat{w}^{b}_{0}(\eta)+ R^{-1} [\hat{w}^{r}_{1}(r)+\hat{w}^{b}_{1}(\eta)]+O\left(R^{-2}\right)  \label{4.12} \\
&& \hat{p}=\hat{p}^{r}_{0}(r)+\hat{p}^{b}_{0}(\eta)+ R^{-1} [\hat{p}^{r}_{1}(r)+\hat{p}^{b}_{1}(\eta)]+O\left(R^{-2}\right)  \label{4.13}
\end{eqnarray}
Here $\eta=R(1-r/a)$ is the boundary layer variable, functions with superscript ``r'' represent
the regular part of the expansion, and functions with superscript ``b'' give us boundary layer corrections
to the regular part. The boundary layer part of the expansion exponentially decays outside the boundary layer.
A brief account of the details of the asymptotic expansion is given in Appendix C. Here we will only present the results.

\begin{figure}
\begin{center}
\includegraphics*[height=7cm]{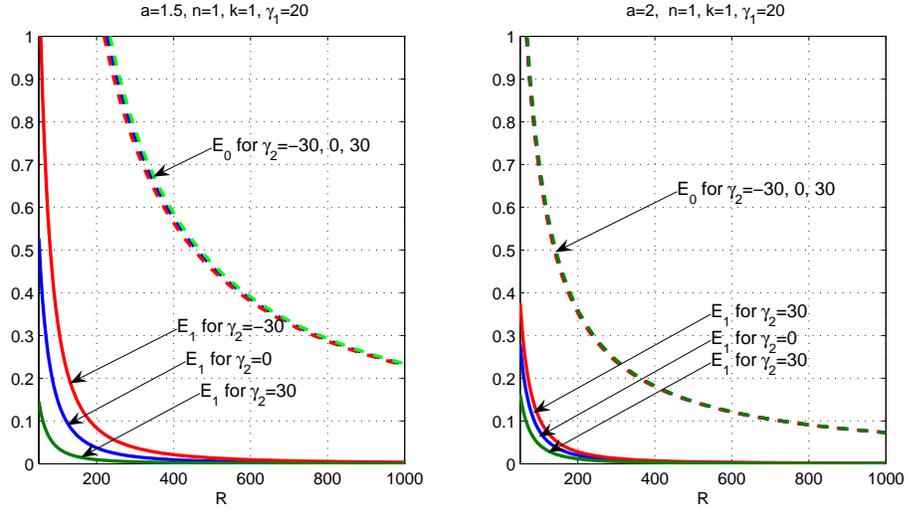}
\end{center}
\caption{$E_0\equiv \vert \sigma-\sigma_0\vert$ and $E_1\equiv \vert \sigma-\sigma_0-\sigma_1/R\vert$ versus $R$.
In both plots, $n=1$, $k=1$ and $\gamma_1=20$. (a) corresponds to $a=1.5$ and (b) shows the results for $a=2$.}
\label{visc1}
\end{figure}

\begin{figure}
\begin{center}
\includegraphics*[height=6cm]{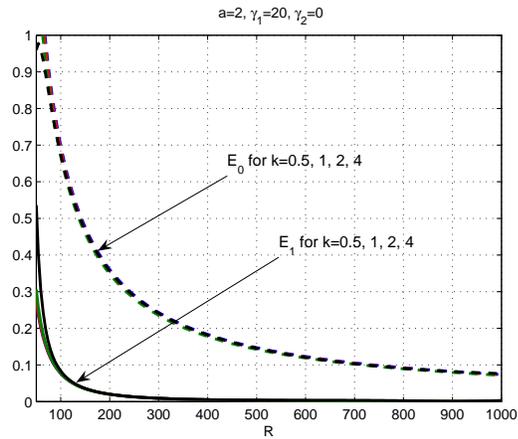}
\end{center}
\caption{$E_0\equiv \vert \sigma-\sigma_0\vert$ and $E_1\equiv \vert \sigma-\sigma_0-\sigma_1/R\vert$ versus $R$:
$a=2$, $n=1$, $\gamma_1=20$, $\gamma_2=0$ and $k=0.5, 1, 2$ and $4$. }
\label{visc2}
\end{figure}

In Eq. (\ref{4.9}), $\sigma_0$ is the inviscid eigenvalue discussed in the previous section, and $\sigma_{1}$ is the first-order
viscous correction, computed in Appendix $C$. The exact eigenvalue problem, given by Eqs. (\ref{4.4})--(\ref{4.8})
was solved numerically using an adapted version of
a Fourier-Chebyshev Petrov-Galerkin spectral method described in \cite{Trefethen2003}.
We have computed the eigenvalue with largest real part, $\sigma$, numerically for a range of values of the Reynolds number $R$ and compared the results with the inviscid eigenvalue $\sigma_0$ and the first order viscous
approximation $\sigma_0+\sigma_1/R$. The
error of approximating $\sigma$ by these is shown in Figs. \ref{visc1} and \ref{visc2} where
$E_0\equiv \vert \sigma-\sigma_0\vert$ and $E_1\equiv \vert \sigma-\sigma_0-\sigma_1/R\vert$.
Figure \ref{visc1} shows the plots of $E_0$ and $E_1$ versus $R$ for $a=1.5$ and $a=2$ for various values of $\gamma_2$.
In both plots, $n=1$, $k=1$ and $\gamma_1=20$. It is evident that in both cases,
$\sigma_0+\sigma_1/R$ gives much
better approximation for $\sigma$ if the Reynolds number is sufficiently high. One can also see that $E_1$ is quite small even for $R=200$,
which is not a very high value of the Reynolds number.
Figure \ref{visc2} shows $E_0$ and $E_1$ as functions of $R$ for $a=2$, $n=1$, $\gamma_1=20$ and $\gamma_2=0$ and for various values of the axial wave number, $k$. One can see that the dependence of $E_0$ and $E_1$ on $k$ is very weak. In fact, all curves shown in Fig. (\ref{visc2}) almost the same as the curve corresponding to two-dimensional perturbation and shown in Fig. 2(b) of Ref. \cite{IM2013b}.

Both Fig. \ref{visc1} and \ref{visc2} show that even for moderate values of $R$ such as $R=200$ the asymptotic formula $\sigma \approx\sigma_0+ \sigma_1/R$ (where $\sigma_1$ is computed in Appendix C) produces
eigenvalues that are very close to the eigenvalues of problem (\ref{4.4})--(\ref{4.8}). This means not only that the inviscid instability studied here persists if viscosity is taken into account, but also that the asymptotic theory works well for Reynolds numbers which are not very high,
and this, in turn, implies that the instability may be observed in real flows.


\section{Discussion}

We have shown that, in the framework of the inviscid theory, a simple rotationally-symmetric flow
between two permeable cylinders is unstable to small three-dimensional perturbations. We gave a rigorous proof
of the facts that the purely radial diverging and converging flows are stable and that unstable modes cannot be axisymmetric.
Numerical calculations demonstrated that (i) for all values of the geometric parameter $a$ in the range from $1.25$ to $8$, the most unstable mode (i.e.
the mode that becomes unstable first when parameter $\gamma$ is increased from $0$) is two-dimensional and (ii) the critical value of $\gamma$
minimized over all azimuthal modes is a strictly increasing function of the axial wave number.

We have also derived an asymptotic expansion of the corresponding viscous
stability problem for $R\gg 1$, computed the first-order viscous correction to inviscid eigenvalues and compared the asymptotic results with
numerically obtained viscous eigenvalues. This demonstrated that the asymptotic results give a very good approximation even for
Reynolds numbers that are not particularly high, such as $R=200$, which suggests that the instability may be be observed in real flows.
Of course, precise conditions under which this instability will be dominating require a further study, and this is a subject
of a continuing investigation.

It is known that a purely azimuthal flow with the velocity inversely proportional to $r$ is stable to three-dimensional perturbations (this follows from a  sufficient condition for stability given in \cite{Howard}). The present paper shows that a purely radial flow is also stable to three-dimensional perturbations. These facts indicate that the physical mechanism of the instability must rely on some destabilising effect arising from the presence of both the radial and azimuthal components of the basic flow. It has been shown in our previous paper \cite{IM2013a} that
if a small radial component is added
to the purely azimuthal flow, it immediately becomes unstable for any value of the ratio of the radii of the cylinders, and the growth rate is proportional to the square root of the ratio of the radial component of the velocity to the azimuthal one.
The asymptotic behaviour of two-dimensional unstable eigenmodes in the limit of weak radial flow (see \cite{IM2013a}) shows that this limit is
a singular limit of the linear stability problem. Adding a weak radial flow to a purely azimuthal one
results in formation of an \emph{inviscid} boundary layer near the inflow part of the boundary, and the unstable
eigenmodes are concentrated within this boundary layer. These facts suggest the following physical mechanism of the instability: in a purely azimuthal flow there are no unstable eigenmodes, but when we add a weak
radial flow, this leads to appearance of new unstable eigenmodes (which do not exist at all if there is no radial flow) concentrated within a thin inviscid boundary layer near the inflow part
of the boundary. This mechanism bears some resemblance to the tearing instability in the magnetohydrodynamics, where
the unstable
tearing mode appears when a small resistivity is taken into consideration (see Ref. \cite{Furth}).

The instability considered here is oscillatory. The two-dimensional neutral modes represent azimuthal travelling waves,
while the three-dimensional ones are helical waves. An oscillatory instability and appearance of azimuthal and helical waves are also
present in the Couette-Taylor flow between impermeable cylinders.
In the Couette-Taylor flow, these waves are observed at moderate azimuthal Reynolds numbers and are associated with viscous effects
(see, e.g., \cite{Iooss}). The results of the present paper show that, in the presence of a radial flow,
azimuthal and helical waves may appear at arbitrarily large radial Reynolds numbers, which means that these waves
can be generated not only by fluid viscosity but also by a radial flow.
This has a certain similarity with self-oscillations observed in numerical simulations of inviscid flows through a channel of finite length
\cite{GMV}. A more detailed analysis of the effect a radial flow on the stability characteristics of the Couette-Taylor flow  requires
a further investigation which would take full account of the viscosity.
A particularly interesting question that arises in this context is
the relation between the instability studied here and the classical
centrifugal instability that leads to the formation of the Taylor vortices.
Here is an interesting paradox: in the inviscid theory, axisymmetric modes cannot be unstable, but it is well known that
the monotonic instability with respect to axisymmetric perturbation occurs in the Couette-Taylor flow with radial flow
(see, e.g., \cite{Min, Serre}).
Our hypothesis is that the monotonic axisymmetric and oscillatory non-axisymmetric instabilities are well separated in the space
of parameters of the problem. If this were so, it would mean that our instability can be observed experimentally.
This, however, requires a further theoretical study and is a topic of a continuing investigation.

The results presented here are mainly of theoretical interest. However, as was argued in Ref. \cite{Gallet2010},
they may be relevant to astrophysical flows such as accretion discs (see also Refs. \cite{Kerswell, Kersale}).
Our results may also shed some light on the
physical mechanism of the formation of strong rotating jets in flows produced by a rotating disk which had been observed experimentally
(see \cite{Petr2001, Petr2002}).

\section*{Acknowledgements}

A. Morgulis acknowledges financial support
from a project which is a part of Russian Government research task No. 1.1398.2014/K.



\section{Appendix A}\label{appA}

Here we will show that, for the diverging flow, (i) there are no unstable modes if the basic flow is purely radial and (ii)
all axisymmetric modes are stable.
To do this, we employ the following theorem of P\'{o}lya (problem 177 of Part V in \cite{Polya_book}, see also \cite{Polya1918}).

\vskip 2mm
\noindent
\textbf{P\'{o}lya's theorem.} {\em Let the function $f(t)$ be continuously differentiable and positive for $0<t<1$, and also let $\int_{0}^{1}f(t)dt$ exist.
The entire function defined by the integral
\[
\int\limits_{0}^{1}f(t)e^{zt}dt=F(z)
\]
has no zeros}
\begin{eqnarray}
&&\hbox{in the half-plane} \quad Re \, z \geq 0, \ \ \hbox{if} \ \ f'(t)>0, \nonumber \\
&&\hbox{in the half-plane} \quad Re \, z \leq 0, \ \ \hbox{if} \ \ f'(t)<0. \nonumber
\end{eqnarray}
It should be noted that the interval $(0,1)$ in the above theorem can be replaced by an arbitrary finite interval $(a,b)$.

Consider first the case of purely radial flow. For $\gamma=0$, the dispersion relation (\ref{3.18}) can be written as
\begin{equation}
D(\sigma)= \frac{1}{a} \, \int\limits_{1}^{a} e^{-\sigma \frac{r^2}{2}} \, \Phi(kr)  \, r \, dr  \label{A1}
\end{equation}
where
\begin{equation}
\Phi(s)=s I_{n}'(s)s_0 K_{n}'(s_0)-s_0 I_{n}'(s_0) s K_{n}'(s), \quad s_{0}\equiv ka .  \label{A2}
\end{equation}
The change of variable of integration, $\xi=r^2/2$, transforms (\ref{A1}) to
\begin{equation}
D(\sigma)= \frac{1}{a} \,  \int\limits_{1/2}^{a^2/2} e^{-\sigma \xi} f(\xi) \, d\xi, \quad f(\xi)\equiv \Phi(k\sqrt{2\xi}).  \label{A3}
\end{equation}
If function $f(\xi)$ were such that $f(\xi)>0$ and $f'(\xi)<0$ for $\xi\in\left(\frac{1}{2},\frac{a^2}{2}\right)$, then the above theorem implies
that $D(\sigma)$ has no zeros in the half-plane $Re \, \sigma \geq 0$, i.e. all its zeros satisfy $Re \, \sigma < 0$, which means that
all modes are stable.

The conditions for function $f(\xi)$ that should be checked are equivalent to the following conditions for $\Phi(s)$:
\begin{equation}
\Phi(s)>0\quad \hbox{and} \quad \Phi'(s)<0 \quad \hbox{for} \ \ s\in\left(k,s_0\right).  \label{A4}
\end{equation}
It is convenient to introduce function $\Psi(s)$ by the formula
\begin{equation}
\Psi(s)=I_{n}(s)s_0 K_{n}'(s_0)-s_0 I_{n}'(s_0) K_{n}(s).  \label{A5}
\end{equation}
Conditions (\ref{A4}), expressed in terms of $\Psi(s)$, become
\begin{eqnarray}
&&s\Psi'(s)>0\quad \hbox{for} \ \ s\in\left(k,s_0\right),  \label{A6} \\
&&\left(s\Psi'\right)'(s)<0\quad \hbox{for} \ \ s\in\left(k,s_0\right).  \label{A7}
\end{eqnarray}
It is easy to see that function $\Psi(s)$ is a solution of the modified Bessel differential equation
\begin{equation}
\frac{1}{s} \, \frac{d}{ds}\left(s \, \frac{d\Psi}{ds}\right)-\left(1+\frac{n^2}{s^2}\right)\Psi=0 \label{A8}
\end{equation}
and satisfies the following boundary conditions:
\begin{equation}
\Psi(s_0)=-1 \quad \hbox{and} \quad \Psi'(s_0)=0, \label{A9}
\end{equation}
where the first of these conditions follows from the Wronskian relation (\ref{3.16}).

First we prove the following auxiliary statement: $\Psi(s)\neq 0$ and $\Psi'(s)\neq 0$ for all $k\leq s < s_0$.
To do this, we assume that either $\Psi(s_{*})=0$ or $\Psi'(s_{*})=0$ for some $s_{*}\in [k,s_0)$. Multiplying Eq. (\ref{A8})
by $s \, \Psi(s)$ and integrating from $s_{*}$ to $s_0$, we find that
\begin{eqnarray}
\int\limits_{s_*}^{s_0}\left(1+\frac{n^2}{s^2}\right)\Psi^2(s) s\, ds &=&
\int\limits_{s_*}^{s_0} \Psi(s)\left(s \, \Psi'\right)'(s) \, ds \nonumber \\
&=& s \, \Psi(s)\Psi'(s)\Bigm\vert_{s_*}^{s_0} - \int\limits_{s_*}^{s_0}\Psi'^{2}(s) s\, ds. \nonumber
\end{eqnarray}
Therefore, if either $\Psi(s_*)=0$ or $\Psi'(s_*)=0$, then
\[
\int\limits_{s_*}^{s_0}\left(1+\frac{n^2}{s^2}\right)\Psi^2(s) s\, ds = - \int\limits_{s_*}^{s_0}\Psi'^{2}(s) s\, ds ,
\]
which is impossible. Therefore,  both $\Psi(s)$ and $\Psi'(s)$ must be nonzero for all $k\leq s < s_0$.

Now we are ready to prove the required properties of $\Psi(s)$. Since $\Psi(s_0)<0$ and $\Psi(s)$ cannot change sign
for $s\in[k,s_0)$, we conclude that $\Psi(s) < 0 $ for all $s\in[k,s_0)$. Then, in view of the differential equation (\ref{A8}),
we obtain
\[
\left(s \, \Psi'\right)'(s)=s \left(1+\frac{n^2}{s^2}\right)\Psi(s) \quad \Rightarrow \quad \left(s \, \Psi'\right)'(s)<0.
\]
We have thus proved (\ref{A7}).
To prove (\ref{A6}), we observe that it follows from the differential equation (\ref{A8}) and the boundary conditions that
$\Psi''(s_0)<0$. This means that $\Psi'(s) > \Psi'(s_0)=0$ at least near the end point $s=s_0$. But since
$\Psi'(s)$ cannot change sign, it must be positive for all $s\in[k,s_0)$, so that condition (\ref{A6}) is satisfied.

Thus, the P\'{o}lya theorem implies that for the purely radial converging basic flow ($\gamma=0$), there are no
unstable modes.

It is easy to see that for the axisymmetric mode ($n=0$) and for any $\gamma$, the dispersion relation (\ref{3.18}) also reduces to
Eq. (\ref{A3}) with $n=0$, so that we may conclude that there are no growing axisymmetric modes.


\section{Appendix B}\label{appB}

Here we construct an asymptotic expansion of the solution to eigenvalue problem (\ref{14})--(\ref{24}) for large axial wave number $k\gg1$. It is convenient to rewrite this problem in a form different from what has been obtained in Section 3.1.1.

Let $H(r)=r\hat{u}(r)$. Then Eq. (\ref{3.5}) can be written as
\begin{equation}
\frac{in}{r}\, H'(r) -\left(k^2 + \frac{n^2}{r^2}\right) G(r)=0 \label{c1}
\end{equation}
where $G=r\hat{v}(r)$. It follows from (\ref{c1}) that $\hat{\omega}_3(r)$, given by Eq. (\ref{20}), can be rewritten in term of $H$ only as
\[
\hat{\omega}_3 = \frac{in}{r} \, \pr_{r}\left( \frac{r}{n^2+k^2r^2} \, H'(r)\right) - \frac{in}{r^2}\, H(r).
\]
Substituting this into Eq. (\ref{16}) and dropping the inessential factor $in$ yields the equation
\begin{equation}
\left(\sigma+\frac{in\gamma}{r^2}+\frac{\beta}{r} \, \pr_{r} \right)
\left[\frac{1}{r} \, \pr_{r}\left( \frac{r}{n^2+k^2r^2} \, H'(r)\right) - \frac{1}{r^2}\, H(r)\right] =0 . \label{c2}
\end{equation}
Equation (\ref{c2}) must be solved subject to boundary conditions (\ref{21})--(\ref{24}) which, in terms of $H$, can be written as
\begin{equation}
H(1)=0, \quad H(a)=0  \label{c3}
\end{equation}
and either
\begin{equation}
H'(1)=0 \label{c4}
\end{equation}
for the diverging flow ($\beta=1$) or
\begin{equation}
H'(a)=0 \label{c5}
\end{equation}
for the converging flow ($\beta=-1$). Equations (\ref{c4}) and (\ref{c5}) follow from the incompressibility condition (\ref{3.4}).

\vskip 3mm
\noindent
\emph{Diverging flow.} Figure \ref{most_unstable_mode} indicates that the azimuthal number of the most unstable mode behaves like
$n\sim k$ for large $k$. Therefore, in order to capture the stability boundary for large $k$, we consider the limit
\[
k\to\infty, \quad n\to\infty, \quad n\sim k.
\]
So, we set $n=\alpha \, k$ in Eq. (\ref{c2}) where $\alpha >0$ and does not depend on $k$. We also assume that
\begin{equation}
\gamma=\tilde{\gamma} \, k \quad \hbox{and} \quad \sigma=-i\tilde{\gamma} \alpha \, k^2 + \tilde{\sigma}\, k \label{c6}
\end{equation}
where $\tilde{\gamma}=O(1)$ and $\tilde{\sigma}=O(1)$ as $k\to\infty$. Incorporating these assumptions into Eq. (\ref{c2}), we get
\begin{equation}
\left[i\tilde{\gamma}\alpha\left(\frac{1}{r^2}-1\right)+\tilde{\sigma}\, \frac{1}{k} +
\frac{1}{k^2} \, \frac{1}{r} \, \pr_{r} \right]
\left[\frac{1}{k^2} \, \frac{1}{r} \, \pr_{r}\left( \frac{r}{\alpha^2 +r^2} \, H'(r)\right) - \frac{1}{r^2}\, H(r)\right] =0 . \label{c7}
\end{equation}
In the limit $k\to\infty$, this equation reduces to
\[
- i\tilde{\gamma}\alpha\left(\frac{1}{r^2}-1\right)\frac{H}{r^2}=0.
\]
This implies that $H(r)$ must be zero everywhere except a thin boundary layer near $r=1$ where the above leading order term
becomes small ($O(k^{-1})$ as $k\to\infty$) and of the same order as some terms which we have discarded.
To treat this boundary layer, we introduce the boundary layer variable $\xi$ such that
\[
r=1+\frac{1}{k} \, \xi
\]
and rewrite Eq. (\ref{c7}) in terms of $\xi$. At leading order, we obtain
\begin{equation}
\left[\tilde{\sigma} -2i\tilde{\gamma}\alpha \, \xi + \pr_{\xi}\right]
\left[\frac{1}{1+\alpha^2} \, H''(\xi) -  H(\xi)\right] =0 . \label{c8}
\end{equation}
Boundary conditions (\ref{c3}), (\ref{c4}), written in terms of $\xi$, take the form
\begin{equation}
H(0)=0, \quad H'(0)=0, \quad H\to 0 \ \ \hbox{as} \ \ \xi\to\infty. \label{c9}
\end{equation}
The solution of Eq. (\ref{c8}), satisfying the first and the last of conditions (\ref{c9}), can be written as
\begin{equation}
H(\xi)=\frac{C_{1}}{2\sqrt{2}}\, \int\limits_{0}^{\infty} e^{-\tilde{\sigma}s+i\tilde{\gamma}\alpha s^2}
\left[e^{-\mu(\alpha)\vert\xi+s\vert}-e^{-\mu(\alpha)\vert\xi-s\vert}\right] ds  \label{c10}
\end{equation}
where $\mu(\alpha)=\sqrt{1+\alpha^2}$ and $C_1$ is an arbitrary constant. Note that formula (\ref{c10}) is valid only for
$\tilde{\sigma}$ satisfying the condition $\Real(\tilde{\sigma}) > 0$. This means that our asymptotic result can only describe
unstable eigenmodes.

Substituting (\ref{c10}) into the second boundary condition (\ref{c9}), we find that the condition of existence of non-trivial solutions of
problem (\ref{c8}), (\ref{c9}) is
\begin{equation}
D_2(\tilde{\sigma}, \tilde{\gamma}, \alpha) \equiv \int\limits_{0}^{\infty} e^{-\tilde{\sigma}s+i\tilde{\gamma}\alpha s^2}
e^{-\mu(\alpha)  s} ds =0. \label{c11}
\end{equation}
Equation (\ref{c11}) represents the dispersion relation for eigenvalues $\tilde{\sigma}$. Note that
$D_2(\tilde{\sigma}, \tilde{\gamma}, \alpha)$, given by this formula,
makes sense for $\tilde{\sigma}$ such that $\Real(\tilde{\sigma}) > -\mu(\alpha)$ and may have zeroes with
$-\mu(\alpha) < \Real(\tilde{\sigma}) \leq 0$. However, only zeros of
$D_2(\tilde{\sigma}, \tilde{\gamma}, \alpha)$ with $\Real(\tilde{\sigma}) > 0$ represent asymptotic approximations to
the eigenvalues of the original problem.
\begin{figure}
\begin{center}
\includegraphics*[height=7cm]{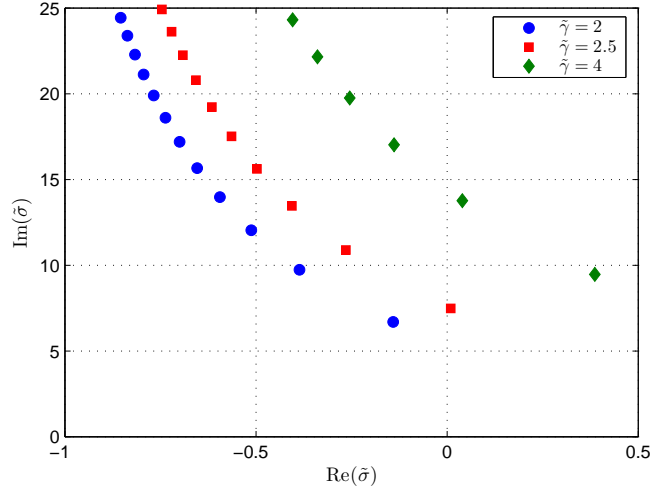}
\end{center}
\caption{Roots of Eq. (\ref{c12}) for $\alpha=1$ and $\tilde{\gamma}=2, 2.5, 4$.}
\label{asymptot_eigs}
\end{figure}

To make calculations easier, it is convenient to transform the dispersion relation to an equivalent form
by deforming the path of integration on the complex plane of variables $s$ from the positive real axis to the half-line:
$s=r\, e^{i\pi/4}$, $r\in[0,\infty)$. Then the dispersion relation takes the form
\begin{equation}
\int\limits_{0}^{\infty} e^{-\tilde{\gamma}\alpha \, r^2-e^{i\pi/4}[\tilde{\sigma}+\mu(\alpha)] \, r} dr =0. \label{c12}
\end{equation}
This equation can be further rewritten in term of the error function $\mathrm{erf}(z)$, but we do not do this here as Eq. (\ref{c12})
is more convenient for numerical calculations.
Typical roots of Eq. (\ref{c12}) are shown in Fig. \ref{asymptot_eigs}. Evidently, when $\tilde{\gamma}(\alpha)$ is smaller than some critical value
$\tilde{\gamma}_{cr}(\alpha)$, all roots are in the left half plane, and
when $\tilde{\gamma}(\alpha) > \tilde{\gamma}_{cr}(\alpha)$, there is at least one root with $\Real(\tilde{\sigma}) > 0$, which represents
an asymptotic approximation of an unstable eigenvalue in the original problem.

To determine $\tilde{\gamma}_{cr}(\alpha)$ and ${\lambda}_{cr}(\alpha)$, we require
$\tilde{\sigma}$ to be purely imaginary, i.e. $\tilde{\sigma}=i\lambda$ with $\lambda\in\mathbb{R}$. The dispersion relation (\ref{c12}) becomes
\begin{equation}
\int\limits_{0}^{\infty} e^{-\tilde{\gamma}\alpha \, r^2-e^{i\pi/4}[i\lambda +\mu(\alpha)] \, r} dr =0. \label{c13}
\end{equation}
Equation (\ref{c13}) has many roots. These roots corresponds to values of $\tilde{\gamma}(\alpha)$ at which
one of the zeros of the function $D_2(\tilde{\sigma}, \tilde{\gamma}, \alpha)$ crosses the imaginary axis on the complex $\tilde{\sigma}$-plane.
We are only interested in the root that corresponds to the smallest value of $\tilde{\gamma}$ because it is this root that
determines the stability boundary. In what follows we will consider only this root of Eq. (\ref{c13}).

It turns out (the proof is below) that the function $\tilde{\gamma}_{cr}(\alpha)$ has a local minimum at $\alpha=1$
(i.e. when $n=k$ as $k\to\infty$). Calculations yield
\[
\min_{\alpha}\tilde{\gamma}_{cr}(\alpha)=\tilde{\gamma}_{cr}(1)\approx 2.4671 \quad \hbox{and} \quad
\lambda_{cr}(1)\approx 7.4331 .
\]
Since we are interested in the asymptotic behaviour of the stability boundary on the $(k,\gamma)$, we choose $\alpha=1$ corresponding to the
minimum value of $\tilde{\gamma}_{cr}(\alpha)$. Thus, the behaviour of the stability boundary in the limit
$k\to\infty$ is given by
\[
\min_{\alpha}{\gamma}_{cr}(\alpha,k)=2.4671 \, k + O(1).
\]
This is shown by circles in Fig. \ref{min_cr_gamma}.

To prove that $\tilde{\gamma}_{cr}(\alpha)$ attains its minimum value at $\alpha=1$,
we change the variable of integration in Eq. (\ref{c13}): $r=\zeta \sqrt{2}/\mu(\alpha)$. This transforms (\ref{c13})
to the equation
\begin{equation}
\frac{\sqrt{2}}{\mu(\alpha)}
\int\limits_{0}^{\infty} e^{-{\gamma}_{*} \, \zeta^2-e^{i\pi/4}[i\lambda_{*} +\sqrt{2}] \, \zeta} d\zeta =0 \label{c13b}
\end{equation}
where
\[
{\gamma}_{*}=\frac{2\alpha}{\mu^2(\alpha)} \, \tilde{\gamma}, \quad \lambda_{*}=\frac{\sqrt{2}}{\mu(\alpha)} \, \lambda.
\]
Then we make an observation that, up to an inessential constant factor, Eq. (\ref{c13b}) is exactly the same as Eq. (\ref{c13}) for $\alpha=1$, with
$\tilde{\gamma}$ and $\lambda$ replaced by ${\gamma}_{*}$ and $\lambda_{*}$. This implies that, if ${\gamma}_{*}$ and $\lambda_{*}$ represent
a root of Eq. (\ref{c13b}), then ${\gamma}_{*}=\tilde{\gamma}_{cr}(1)$ and ${\lambda}_{*}={\lambda}_{cr}(1)$. This fact and the definition of ${\gamma}_{*}$ and $\lambda_{*}$ have a consequence that
\[
\tilde{\gamma}_{cr}(\alpha) = \frac{1+\alpha^2}{2\alpha} \, \tilde{\gamma}_{cr}(1), \quad
\lambda_{cr}(\alpha)= \frac{\sqrt{1+\alpha^2}}{\sqrt{2}} \, {\lambda}_{cr}(1).
\]
Since function $f(\alpha)=(1+\alpha^2)/(2\alpha)$ attains its minimum value at $\alpha=1$ and $f(1)=1$, we obtain the required property of
$\tilde{\gamma}_{cr}(\alpha)$.

\vskip 3mm
\noindent
\emph{Converging flow.} For the converging flow, a similar analysis shows that in the limit
\[
ka\to\infty, \quad n=\alpha \, ka,
\]
the eigenvalue problem (\ref{c2}), (\ref{c3}), (\ref{c5}) reduces to
\begin{equation}
\left[\tilde{\sigma} +2i\tilde{\gamma}\alpha \, \eta + \pr_{\eta}\right]
\left[\frac{1}{1+\alpha^2} \, H''(\eta) -  H(\eta)\right] =0  \label{c14}
\end{equation}
and
\begin{equation}
H(0)=0, \quad H'(0)=0, \quad H\to 0 \ \ \hbox{as} \ \ \eta\to\infty. \label{c15}
\end{equation}
where $\eta$, $\tilde{\gamma}$ and $\tilde{\sigma}$ are defined by
\[
\eta=ka\left(1-\frac{r}{a}\right), \quad \gamma=\tilde{\gamma} \, ka, \quad
\sigma=\frac{1}{a^2} \, \left[-i\tilde{\gamma}\alpha(ka)^2+ \tilde{\sigma} \, ka\right].
\]
It is easy to see that, the complex conjugate of Eq. (\ref{c14}) is equivalent to Eq. (\ref{c8}),
with $\tilde{\sigma}$ replaced by its complex conjugate $\overline{\tilde{\sigma}}$. This means
that if $\tilde{\sigma}$ and $H(\xi)$ represent a solution of problem (\ref{c8}), (\ref{c9}), then $\overline{\tilde{\sigma}}$ and
and $\overline{H}(\eta)$ solve problem (\ref{c14}), (\ref{c15}). Therefore, the asymptotic result for the converging flow
can be obtained from that for the diverging flow by simply replacing $\tilde{\sigma}$ by $\overline{\tilde{\sigma}}$ and
$H(\xi)$ by $\overline{H}(\eta)$. Hence, we obtain
\[
\min_{\alpha}\tilde{\gamma}_{cr}(\alpha)=\tilde{\gamma}_{cr}(1)\approx 2.4671, \quad  \lambda_{cr}(1)\approx -7.4331
\]
where $\lambda_{cr}=\Imag(\tilde{\sigma})$ when $\Real(\tilde{\sigma})=0$.
This means that in the limit $ka\to\infty$, $n=\alpha \, ka$, we have
\[
\min_{\alpha}{\gamma}_{cr}(\alpha,ka)=2.4671 \, ka + O(1).
\]


\section{Appendix C}\label{appC}

Here we derive the asymptotic approximation (\ref{4.9})--(\ref{4.13}). To obtain the regular part of the expansion (that is
valid everywhere except the boundary layer near $r=a$), we substitute the asymptotic formula for the azimuthal velocity (\ref{4.2})
and Eqs. (\ref{4.9})--(\ref{4.13}) into (\ref{4.4})--(\ref{4.7}), discard all boundary layer terms and collect term containing equal powers of
$1/R$. As a result, we obtain a sequence of equations, the first two of which can be written as
\begin{eqnarray}
&&K\mathbf{v}_{0}^{r}=0, \label{C1} \\
&&K\mathbf{v}_{1}^{r}= - \sigma_1 \mathbf{v}_{0}^{r} + B\mathbf{v}_{0}^{r}. \label{C2}
\end{eqnarray}
Here $\mathbf{v}_{k}^{r}=(\hat{u}_{k}^{r}, \hat{v}_{k}^{r}, \hat{w}_{k}^{r})$ for $k=0,1$ and operators $K$ and $B$ are defined as
\begin{eqnarray}
&&K\mathbf{v}_{k}^{r}=
\left(
\begin{array}{l}
\left(\frac{h'(r)}{r} +  \frac{1}{r} \, \pr_r \right) \hat{u}_{k}^{r}-
\frac{1}{r^2} \, \hat{u}_{k}^{r} -\frac{2\gamma_1}{r^2} \, \hat{v}_{k}^{r} + \pr_r \, \hat{p}_{k}^{r} \\
\left(\frac{h'(r)}{r} +  \frac{1}{r} \, \pr_r \right) \hat{v}_{k}^{r} + \frac{1}{r^2} \, \hat{v}_{k}^{r} + \frac{in}{r} \, \hat{p}_{k}^{r} \\
\left(\frac{h'(r)}{r} +  \frac{1}{r} \, \pr_r \right) \hat{w}_{k}^{r}  + ik \, \hat{p}_{k}^{r}
\end{array}
\right)
\label{C3}
\end{eqnarray}
and
\begin{eqnarray}
&&B\mathbf{v}_{0}^{r}=
\left(
\begin{array}{l}
L \hat{u}_{0}^{r}- \frac{1}{r^2} \, \hat{u}_{0}^{r} - \frac{2in}{r^2} \, \hat{v}_{0}^{r}  \\
L \hat{v}_{0}^{r} - \frac{1}{r^2} \, \hat{v}_{0}^{r} + \frac{2in}{r^2} \, \hat{u}_{0}^{r} \\
L \hat{w}_{0}^{r}
\end{array}
\right).
\label{C4}
\end{eqnarray}
In Eq. (\ref{C3}),
\[
h(r)=\sigma_0 \, \frac{r^2}{2} + in\gamma_{1}\log r.
\]
and $\hat{p}_{k}^{r}$ can be eliminated using the incompressibility condition
\begin{equation}
\pr_r \left(r \hat{u}_{0}^{r}\right) +in \, \hat{v}_{0}^{r} + ik r \hat{w}_{0}^{r}=0. \label{C5}
\end{equation}
Boundary conditions for $\mathbf{v}_{0}^{r}$ and $\mathbf{v}_{1}^{r}$ are obtained by substituting (\ref{4.10})--(\ref{4.12}) into
(\ref{4.8}) and collecting terms containing equal powers of $1/R$. This yields
\begin{eqnarray}
&&\hat{u}_{0}^{r}(1)=\hat{u}_{0}^{r}(a)=0, \label{C6} \\
&&\hat{v}_{0}^{r}(1)=\hat{w}_{0}^{r}(1)=0, \label{C7} \\
&&\hat{v}_{0}^{r}(a)+\hat{v}_{0}^{b}(0)=0, \quad \hat{w}_{0}^{r}(a)+\hat{w}_{0}^{b}(0)=0, \label{C8} \\
&&\hat{u}_{1}^{r}(1)=0, \quad \hat{u}_{1}^{r}(a)+\hat{u}_{0}^{b}(0)=0, \label{C9} \\
&&\hat{v}_{1}^{r}(1)=\hat{w}_{1}^{r}(1)=0 . \label{C10}
\end{eqnarray}
We did not present boundary conditions for $\hat{v}_{1}^{r}$ and $\hat{w}_{1}^{r}$ at $r=a$, as they are not needed in what follows.

Equation (\ref{C1}) and boundary conditions (\ref{C6}) and (\ref{C7}) represent the inviscid eigenvalue problem that was considered
in Section 3. After it is solved, we know $\sigma_0$ and $\mathbf{v}_{0}^{r}$. Then boundary conditions (\ref{C8}) are used to find
the boundary layer corrections $\hat{v}_{0}^{b}$ and $\hat{w}_{0}^{b}$. After that, the boundary layer part of the radial component
of the velocity, $\hat{u}_{0}^{b}$, can be found from the incompressibility condition. Once, $\hat{u}_{0}^{b}$ is known, Eqs. (\ref{C9})
and (\ref{C10}) give us boundary conditions for Eq. (\ref{C2}), which can then be solved, and the entire procedure can repeated as many times
as necessary yielding higher order approximations. However, to find $\sigma_1$, we do not need to calculate the solution of (\ref{C2}) expicitly,
all we need is to ensure that a solution does exist.
Before describing how this can be done, we need to say a few words about the boundary layer.

The boundary layer approximations are obtained as follows. We substitute Eqs. (\ref{4.2})
and (\ref{4.9})--(\ref{4.13}) into (\ref{4.4})--(\ref{4.7}) and take into account that the regular part satisfies Eqs. (\ref{C1}) and (\ref{C2}).
Then we make the change of variable
$r=a(1-R^{-1} \, \eta)$, expand every function of $a(1-R^{-1} \, \eta)$ in Taylor's series at $R^{-1}=0$
and, finally, collect terms of the equal powers of $R^{-1}$. At leading order, the boundary layer equations are given by
\begin{eqnarray}
&&\pr_{\eta}^2\hat{v}_{0}^{b}+\pr_{\eta}\hat{v}_{0}^{b}=0, \label{C11} \\
&&\pr_{\eta}^2\hat{w}_{0}^{b}+\pr_{\eta}\hat{w}_{0}^{b}=0, \label{C12} \\
&&-\pr_{\eta}\hat{u}_{0}^{b} + in\hat{v}_{0}^{b} + ik a\hat{w}_{0}^{b}=0. \label{C13}
\end{eqnarray}
The solutions of Eqs. (\ref{C11})--(\ref{C13}) that satisfy boundary conditions (\ref{C8}) and the condition of decay at infinity
are
\[
\hat{v}_{0}^{b}=-\hat{v}^{r}_{0}(a) \, e^{-\eta}, \quad \hat{w}_{0}^{b}=-\hat{w}^{r}_{0}(a) \, e^{-\eta}, \quad
\hat{u}_{0}^{b}=- \int\limits_{\eta}^{\infty} \left(in\hat{v}_{0}^{b}(s) + ik a\hat{w}_{0}^{b}(s)\right)ds .
\]
Here the constant of integration has been chosen so as to guarantee that $\hat{u}_{0}^{b}(\eta)$ decays at infinity.
Hence, the boundary condition for $\hat{u}_{1}^{r}(r)$ at $r=a$ can be written as
\begin{equation}
\hat{u}_{1}^{r}(a)= - \hat{u}_{0}^{b}(0)= - in\hat{v}_{0}^{r}(a) - ik a\hat{w}_{0}^{i}(a). \label{C14}
\end{equation}

Now consider the non-homogeneous equation (\ref{C2}). It has a solution only if its right hand side satisfies a certain
solvability condition. To formulate it, we define
the inner product
\[
\langle \mathbf{g}, \mathbf{f}\rangle = \int\limits_{1}^{a}\overline{\mathbf{g}}\cdot \mathbf{f} \, r \, dr
= \int\limits_{1}^{a}\left(\overline{g}_{1}f_{1}+\overline{g}_{2}f_{2}+\overline{g}_{3}f_{3}\right) \, r  dr
\]
where $\mathbf{g}=(g_1,g_2,g_3)$, $\mathbf{f}=(f_1,f_2,f_3)$, and $\overline{\mathbf{g}}$ is the complex conjugate of ${\mathbf{g}}$.
With respect to this inner product, we define the adjoint operator $\mathbf{g}\mapsto K^* \mathbf{g}$ by
\[
\langle \mathbf{g}, K\mathbf{f}\rangle = \langle K^*\mathbf{g}, \mathbf{f}\rangle
\]
for any functions $\mathbf{f}$ and $\mathbf{g}$ satisfying the incompressibility conditions
\[
\pr_r \left(r f_1\right) +in \, f_2 + ik r f_3=0, \quad
\pr_r \left(r g_1\right) +in \, g_2 + ik r g_3=0
\]
and the boundary conditions
\begin{eqnarray}
&&f_1(1)=f_2(1)=f_3(1)=0, \quad f_1(a)=0, \label{C15} \\
&&g_1(1)=0, \quad g_1(a)=g_2(a)=g_3(a)=0. \label{C16}
\end{eqnarray}
Note that the boundary conditions for $\mathbf{f}$ and $\mathbf{g}$ are different.

Now let $\mathbf{g}$ satisfy boundary conditions (\ref{C16}) and be a solution of the equation
\[
K^*\mathbf{g} =
\left(
\begin{array}{l}
\left(\frac{\overline{h}'(r)}{r} -  \frac{1}{r} \, \pr_r \right) g_{1}-
\frac{1}{r^2} \, g_1  + \pr_r \, \alpha \\
\left(\frac{\overline{h}'(r)}{r} -  \frac{1}{r} \, \pr_r \right) g_2 + \frac{1}{r^2} \, g_2
- \frac{2\gamma_1}{r^2} \, g_1+ \frac{in}{r} \, \alpha \\
\left(\frac{\overline{h}'(r)}{r} -  \frac{1}{r} \, \pr_r \right) g_3 + ik \, \alpha
\end{array}
\right) = 0
\]
where function $\alpha$ can be eliminated using the incompressibility condition for $\mathbf{g}$.
Taking inner product of Eq. (\ref{C2}) with $\mathbf{g}$ gives us the required
solvability condition:
\[
Q_1= - \sigma_1 Q_2 + Q_3
\]
where
\[
Q_1=\langle \mathbf{g}, K \mathbf{v}_{1}^{r}\rangle, \quad
Q_2= \langle \mathbf{g}, \mathbf{v}_{0}^{r}\rangle, \quad
Q_3= \langle \mathbf{g}, B \mathbf{v}_{0}^{r}\rangle.
\]
Hence,
\begin{equation}
\sigma_{1}=\frac{Q_3 - Q_1}{Q_2}.
\end{equation}
Below are the explicit formulae for $Q_1$, $Q_2$ and $Q_3$ that can be obtained after lengthy but standard calculations:
\begin{eqnarray}
&&Q_1= \left(n^2+k^2a^2\right) \, \int\limits_{1}^{a}
e^{-h(r)}\Theta(kr) \, r dr, \nonumber \\
&&Q_2=- \frac{1}{2} \, \int\limits_{1}^{a}
e^{-h(r)}\Phi(kr) \,  r^3 dr, \nonumber \\
&&Q_3= \left\{ \int\limits_{1}^{a}
e^{-h(r)}W(r)\Phi(kr)\, r dr
+ \int\limits_{1}^{a} e^{-h(r)}\left(r^2-1\right)\Psi(kr) \, r dr \, \right\}, \nonumber
\end{eqnarray}
where functions $\Phi(s)$ and $\Psi(s)$ are defined in Appendix A and
\begin{eqnarray}
&&\Theta(s)=I_n(s_0) s K'_n(s)- s I'_n(s) K_n(s_0) \quad (s_{0}=ka), \nonumber \\
&&W(r)= -\sigma_0^2 \, \frac{r^4}{4} + \left(\frac{k^2}{2}+\sigma_0(1-in\gamma_1)\right) r^2
+n^2\left(1+\gamma_1^2\right) \log r . \nonumber
\end{eqnarray}





\end{document}